\documentclass[a4paper]{article}

\usepackage{INTERSPEECH2019}

\usepackage{xcolor}

\title{Context-aware Goodness of Pronunciation for \\ Computer-Assisted Pronunciation Training}
\name{Jiatong Shi$^1$, Nan Huo$^1$, Qin Jin$^2$$^{\ast}$\thanks{$^{\ast}$Corresponding Author.}}
\address{
  $^1$Department of Computer Science, Johns Hopkins University, U.S.A.\\
  $^2$School of Information, Renmin University of China, P.R.China}
\email{\{jiatong\_shi, nhuo1\}@jhu.edu, qjin@ruc.edu.cn}

\begin{document}

\maketitle
\begin{abstract}
  Mispronunciation detection is an essential component of the Computer-Assisted Pronunciation Training (CAPT) systems. State-of-the-art mispronunciation detection models use Deep Neural Networks (DNN) for acoustic modeling, and a Goodness of Pronunciation (GOP) based algorithm for pronunciation scoring. 
  However, GOP based scoring models have two major limitations: i.e., (i) They depend on forced alignment which splits the speech into phonetic segments and independently use them for scoring, which neglects the transitions between phonemes within the segment; 
  (ii) They only focus on phonetic segments, which fails to consider the context effects across phonemes (such as liaison, omission, incomplete plosive sound, etc.). 
  In this work, we propose the Context-aware Goodness of Pronunciation (CaGOP) scoring model. Particularly, two factors namely the transition factor and the duration factor are injected into  CaGOP scoring.
  The transition factor identifies the transitions between phonemes and applies them to weight the frame-wise GOP. Moreover, a self-attention based phonetic duration modeling is proposed to introduce the duration factor into the scoring model. 
  The proposed scoring model significantly outperforms baselines, achieving 20\% and 12\% relative improvement over the GOP model on the phoneme-level and sentence-level mispronunciation detection respectively. 
\end{abstract}

\section{Introduction}

Computer-Assisted Pronunciation Training (CAPT) is an important technology that offers automatic feedback to help users learn new spoken languages \cite{neri2002pedagogy}. Because of its objectiveness, some standardized examinations also use the CAPT system for automatic speech proficiency evaluation (e.g., TOFEL \cite{evanini2013automated}, AZELLA \cite{metallinou2014using}).

\begin{figure}[t]
  \centering
  \includegraphics[width=0.8\linewidth]{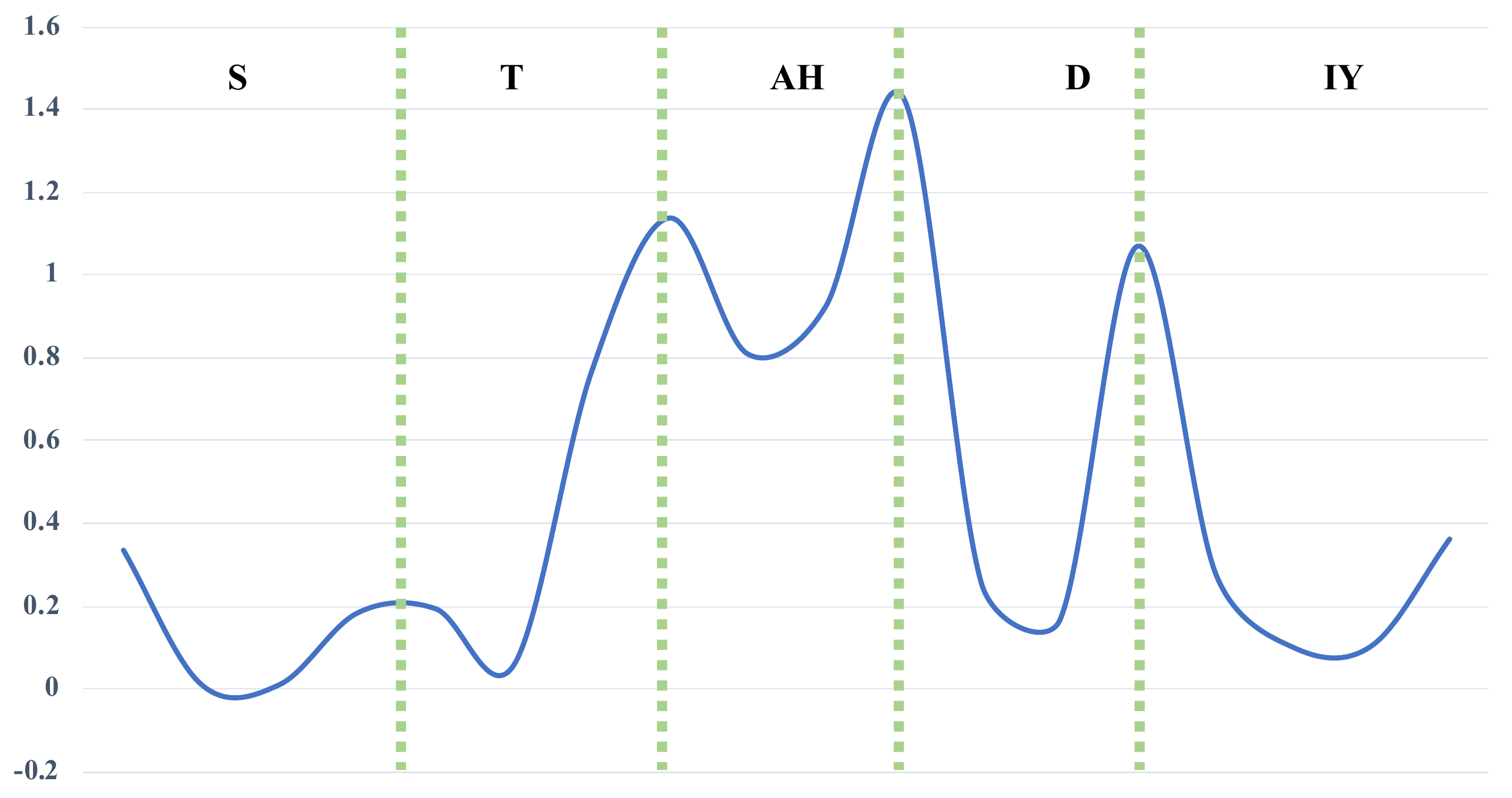}
  \caption{Posterior-Probability Entropy of the word "Study"}
  \label{fig:study-entropy}
\end{figure}

Several directions have been explored for the CAPT problem in the literature. One is to recognize pronunciation error patterns. The patterns can either be pre-defined based on linguistic knowledge or obtained in driven-data way \cite{xu2009automatic, wang2015supervised}. However, it is often hard to implement due to the difficulties of obtaining comprehensive expert knowledge and collecting enough erroneous data. The dominant CAPT systems are based on an Automatic-speech-recognition (ASR)-like architecture \cite{witt2012automatic}. It mainly includes three components: an acoustic 
module, a decoding module, and a scoring module. The acoustic module first converts speech information into frame-level phonetic posterior-probabilities. Next, the decoding module force-aligns the posterior-probabilities into phonetic segments. Lastly, the scoring module scores the given segment according to the reference phoneme. The acoustic module follows the ASR's evolution from Hidden Markov Model-Gaussian Mixture Model (HMM-GMM) \cite{witt1997computer, tsubota2002recognition} to Deep Neural Networks (DNN) \cite{hu2013new, qian2012use, li2016mispronunciation} and Recurrent Neural Networks (RNN) \cite{qian2017bidirectional}. For the scoring module, the dominant method is the Goodness of Pronunciation (GOP) \cite{witt1997computer}. It is a confidence measure for phonetic pronunciation, relating the test speech to an ASR model trained on native speech. Later, weighted GOP (wGOP) was proposed to improve the bad cases when the system cannot successfully identify phonemes with similar sound \cite{van2010using}. Similarly, Zhang et al. employed a confused phoneme set to solve the same problem \cite{zhang2012exploit}.

However, the above GOP-like methods do not fully consider the context information within and between the phonetic segments. Within the segments, the scoring strategies depend on the forced-alignments. The forced-alignments split the entire speech sequence into phonetic segments corresponding to reference phonemes. Based on the facts of speech production, the vocal tract is gradually changing in the production of different phonemes \cite{fant2012auditory}. Therefore, a hard assignment of phonemes in time domain would include the transition between phonemes within the force-aligned segments. As stated in the information theory, the entropy represents the "disorder" degree \cite{shannon2001mathematical}. It reaches the highest for uniform distribution but becomes zero when there is a definite event. The Figure \ref{fig:study-entropy} is an example of the posterior probability entropy for word "Study"\footnote{the posterior-probability is computed from an ASR acoustic model trained from Librispeech \cite{panayotov2015librispeech}.} where the green line stands for the boundary of forced-alignment. We can observe that the entropy rises between phonemes, which indicates that the model has lower confidence level in its predictions. As GOP uses the whole segment for scoring, it may include misleading information (i.e. phonetic transitions) that does not relate to the target phoneme. In addition, the GOP scoring module only considers very short context information using context-dependent phonemes \cite{dahl2011context}.  However, it does not consider longer dependencies (e.g. word-level context information), which may influence the original sound of phonemes (speech co-articulation). As a result, the GOP scoring module tends to give over-strict suggestions when there are accepted co-articulation effects\footnote{For example, they often give a lower score on "D" in "AND" where the "D" is often less-stressed in spoken English} \cite{tong2016context}. 

In this paper, we propose a Context-aware GOP (CaGOP) scoring module which takes into account the transition factor and the duration factor. 
The transition factor is captured using a proposed transition-aware mechanism, while the duration factor depends on a duration prediction model with self-attention. 
On the mispronunciation detection task, the proposed CaGOP outperforms the GOP with around 20\% relative improvement on both accuracy and F1 score. Meanwhile, the sentence-level correlation with human also achieves 12\% relative improvement over the baselines. We also observe 25\% relative improvement in phonetic duration modeling with self-attention structure over the baseline Long-Short-Term-Memory (LSTM) model.

\section{Context-aware GOP}
\begin{figure}[t]
  \centering
  \includegraphics[width=0.8\linewidth]{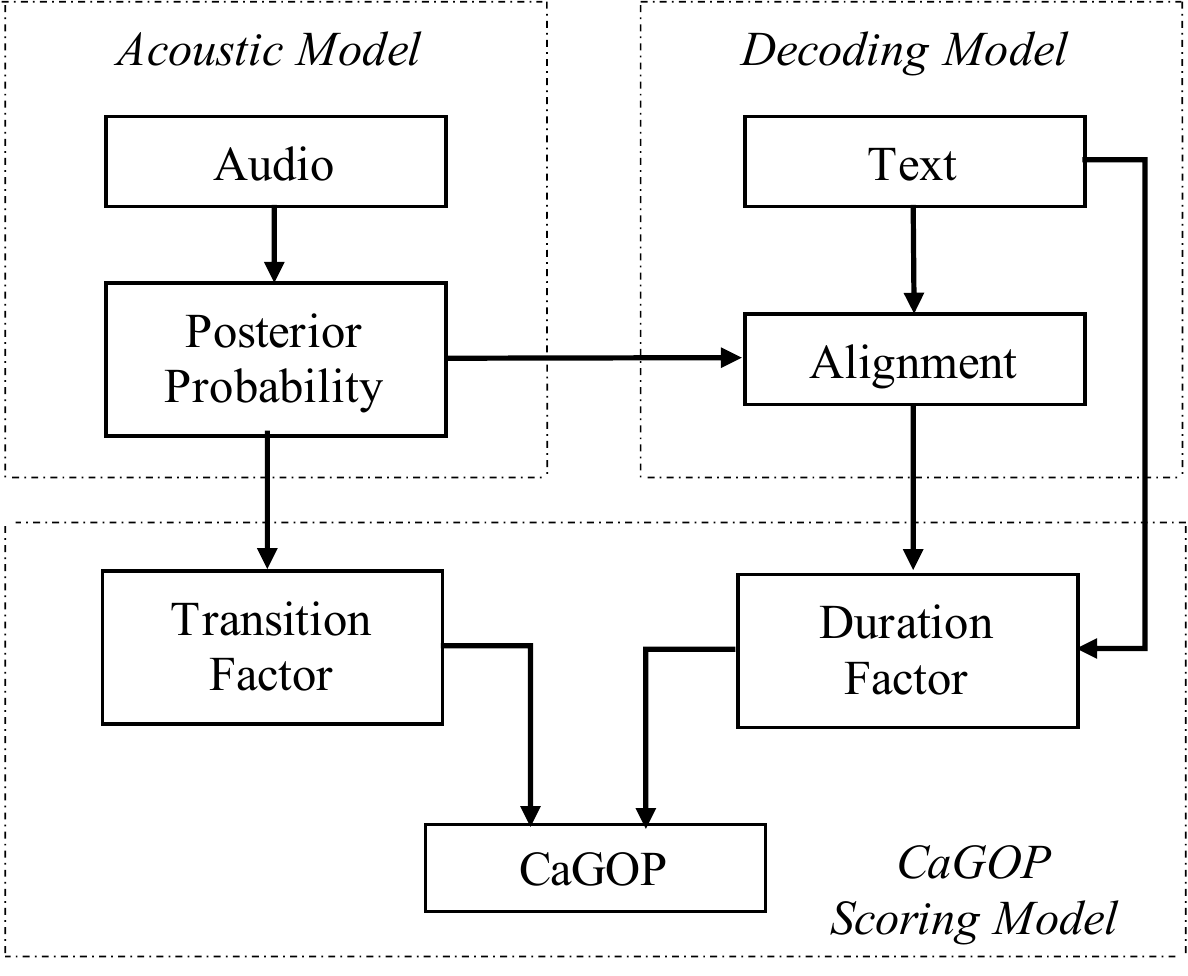}
  \caption{Our proposed CAPT System Framework}
  \label{fig:framework}
\end{figure}

Figure \ref{fig:framework} shows the framework of our CAPT system. Our focus in this paper is on the scoring module. In this section, we first present the background of the GOP and then introduce the transition factor and the duration factor in our proposed context-aware GOP (CaGOP) scoring module.
\vspace{-4pt}
\subsection{Goodness of Pronunciation}
The Goodness of Pronunciation (GOP) scoring module assumes that the orthographic phoneme sequence is given. It applies an acoustic module (e.g. HMM-DNN) to infer the likelihood $p(o_1, o_2, ..., o_N|a')$ of the acoustic feature $o_1, o_2, ... o_N$ corresponding to each phone $a'$ in the phoneme sequence. The final score is defined as the duration normalized log-posterior-probability $\log(p(a|o_1, o_2, ..., o_N))$ for the reference phoneme $a$ \cite{witt1997computer}. Using the Bayesian theorem, it is computed as in Eq. \eqref{eq:gop_definition}.
\vspace{-4pt}
\begin{equation}
    \label{eq:gop_definition}
    \text{GOP(}a\text{)} = \frac{1}{N} \cdot \log(\frac{p(o_1, o_2, ..., o_N|a) \cdot p(a)}{\sum_{a'\in A} p(o_1, o_2, ..., o_N|a') p(a')})
\end{equation}
\noindent
where $a$ stands for the phone index to be scored, $A$ stands for the phone set, and the $N$ represents the length of the acoustic features. Practically, all the phones are assumed to be equal-likely (i.e., $p(a)=p(a')$). Hence, GOP can be approximated into Eq. \eqref{eq:gop_definition_next}. Due to the logarithmic operation, GOP is not bounded to a certain range. The mispronunciation is determined by phone-dependent thresholds.
\begin{equation}
    \label{eq:gop_definition_next}
    \text{GOP(}a\text{)} = \frac{1}{N} \cdot \log(\frac{p(o_1, o_2, ..., o_N|a)}{\sum_{a'\in A} p(o_1, o_2, ..., o_N|a')})
\end{equation}

\subsection{Transition Factor}
As mentioned above, using the whole segments from forced alignment should consider the phone transitions. Inspired by the entropy concept in information theory which refers to the uncertainty of random variables \cite{shannon2001mathematical} and the Entropy Weight Method which is defined based on the entropy to measure the contract intensity of each attribute \cite{zeleny2012linear}, we propose to calculate the entropy weight for frame-level posterior-probabilities in order to reduce the transition effect on the scoring.

For each frame, a lower entropy of posterior-probability represents that the acoustic module is more confident in the prediction, while the fluctuation in posterior-probability entropy indicates that there probably exists phonetic transition within the phonetic segments. 
In order to incorporate the transition factor into the scoring, we first reformulate the GOP to frame-wise GOP using conditional independence \footnote{the conditional independence is inferred from HMM assumptions for the acoustic module} as Eq. \eqref{eq:frame-gop} 
and the frame-wise posterior-probability entropy defined as Eq. \eqref{eq:entropy}. 
\begin{equation}
    \label{eq:frame-gop}
    \text{GOP(}a\text{)} = \frac{1}{N}\sum_{t=1}^N \log p(a|o_t) = \frac{1}{N}\sum_{t=1}^N \log(\frac{p(o_t|a)}{\sum_{a'\in A}p(o_t|a')} )
\end{equation}

\begin{equation}
    \label{eq:entropy}
    E_{t} = -\sum_{a' \in A} p(a'|o_t) \cdot \log(p(a'|o_t))
\end{equation}
We then compute the transition-aware pronunciation score by weighting the frame-wise GOP with the reciprocal of entropy:
\begin{equation}
    \label{eq:transition-aware-gop}
    \text{TAScore}(a) = \sum_t^N \frac{\frac{1}{E_t}}{\sum_{t'}^N \frac{1}{E_{t'}}} \cdot \log(p(a|o_t))
\end{equation}

\subsection{Duration Factor}
As the GOP based model does not consider the context information between phonetic segments, it does not consider the pronunciation phenomena from contextual phonemes. To tackle this issue, we introduce duration factor as the prosody context to the pronunciation scoring. The computation of the duration factor includes two steps. First, we use a context-dependent duration model to predict the duration for the given phoneme sequences. Then, we refer the duration factor as the phonetic duration mismatch between the reference and the test utterances. 

\begin{figure}[t]
  \centering
  \includegraphics[width=0.8\linewidth]{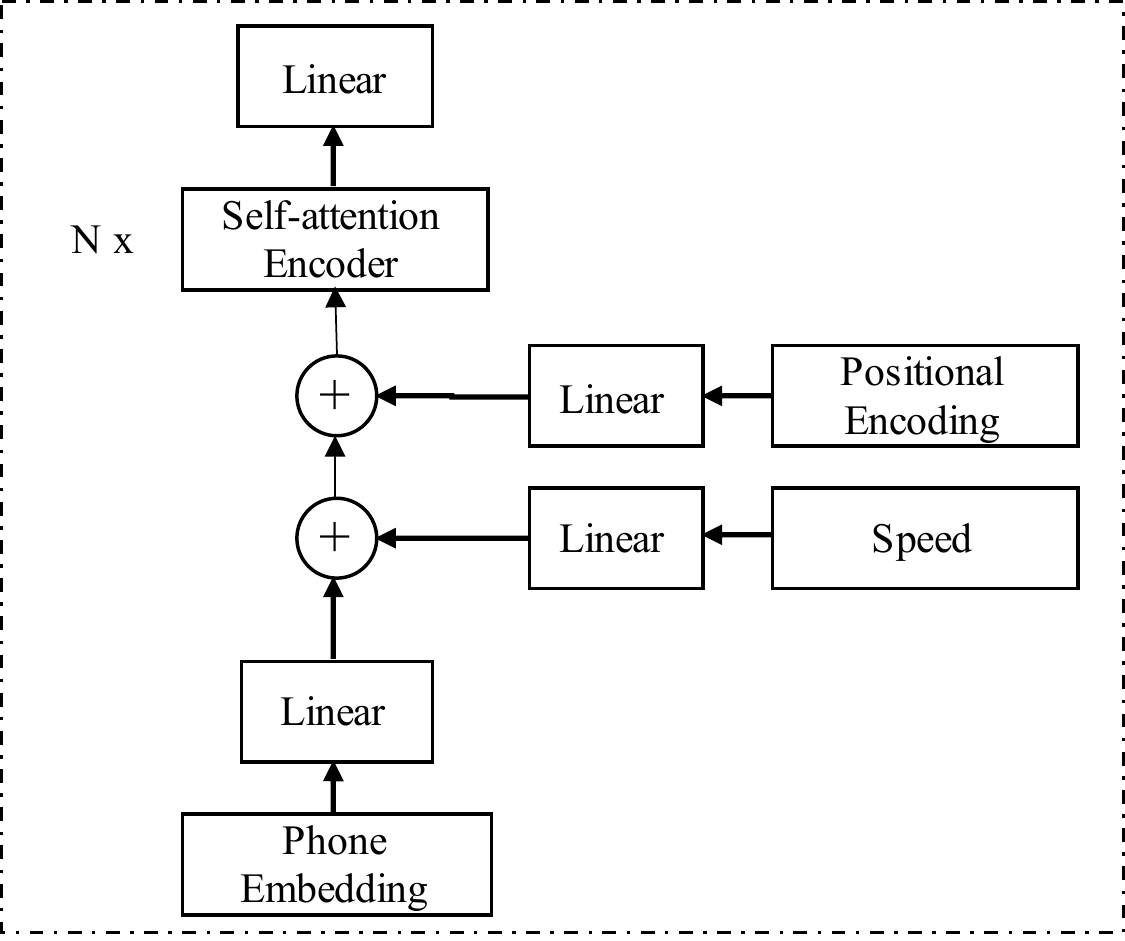}
  \caption{Phonetic Duration Prediction with Self-attention}
  \label{fig:duration-frame-work}
\end{figure}

Phonetic duration is an essential factor in speech articulation and benefits many speech processing tasks (e.g., Text-to-speech, ASR) \cite{pylkkonen2004duration, tokuda2016temporal}. Previous works have explored statistical graphic models (e.g., HMM) \cite{pylkkonen2004duration} and neural networks \cite{henter2016robust, chen2017discrete, wei2019neural}. In this work, we propose to use the multi-head self-attention structure to model the phonetic duration for a given phoneme sequence. The model structure is shown in Figure \ref{fig:duration-frame-work}. First, we pass the reference text to the text-to-phone converter and use phoneme index sequences as our model input. Then we add the speed and positional information. The speed is represented as the average duration of phonemes in the given speech. The self-attention encoder follows the definition of \cite{vaswani2017attention} as shown in Eq. \eqref{eq:attention} but with a local diagonal Gaussian matrix \cite{sperber2018self}. We introduce the diagonal Gaussian matrix to favor local information because phonetic duration does not have very long dependency. The co-articulation and omission of phonemes are usually caused by nearby phonemes.
\vspace{-6pt}
\begin{equation}
    \label{eq:attention}
    \text{Attention}(\mathbf{q}, \mathbf{k}, \mathbf{v}) = \text{softmax}(\frac{\mathbf{q}\cdot \mathbf{k}^T}{\sqrt{d }}+ \mathbf{M}) \cdot \mathbf{v} 
\end{equation}
\vspace{-6pt}
\begin{equation}
    \label{eq:matrix}
    \mathbf{M}_{j, k} = -\frac{(j-k)^2}{\sigma^2}
\end{equation}
where $d$ is the dimension of the input vector and $\mathbf{M}$ is the local Gaussian matrix. The $\mathbf{M}$ in Eq. \eqref{eq:attention} is defined as in Eq \eqref{eq:matrix}, which is a $T \times T$ matrix for the sequence length of $T$, and $\sigma$ is a learnable parameter of the network.

We compute the duration mismatch between the reference and the test utterances as the duration factor. As the phone duration strongly correlates with the speed of speech, we apply phone-speed dependent balance factors to compute the duration mismatch. The factors are computed using the phonetic duration of the training set. For each sentence speed, we compute the absolute error between our reference duration prediction and the ground truth \footnote{we generate the ground truth duration using alignments from the ASR model}. The factors $T_{a, s}$ are introduced as the summation of mean and 1.5 times of the standard deviation of the absolute error. Our estimation of the duration mismatch is defined as Eq. \eqref{eq:prosodic_scoring}.
\begin{equation}
    \label{eq:prosodic_scoring}
    \delta (a) = |D_{align} - D_{pred}| - T_{a, s}
\end{equation}
where $D_{align}$ is the force-aligned duration of phone $a$ (i.e., the ground truth), $D_{pred}$ is the prediction from the reference duration network of phone $a$, and $T_{a, s}$ is the balance factor of phone $a$ and speed $s$. Noted that the $\delta(a)$ can be negative, which indicates that the duration of the target phoneme is as expected.

\vspace{-4pt}
\subsection{CaGOP Scoring}
Injecting both the transition factor and duration factor into the pronunciation scoring, we compute the context-aware pronunciation score as follows:
\begin{equation}
    \label{eq:CaGOP-compute}
    \text{CaGOP}(a) = (1 - \beta \cdot \delta(a)) \cdot \text{TAScore}(a)
\end{equation}
where $\beta$ is the hyper-parameter. Please note that replacing $\text{TAScore}(a)$ with $\text{GOP}(a)$ will lead to a reduced model with only injecting the duration factor, which will be compared with in the following experiments. 

\section{Experimental Settings}
In this section, we first introduce the dataset. Then we present the implementation details for the proposed system and baselines. Since the main focus of this work is on the scoring module, we only briefly describe the settings of the acoustic and decoding modules. Finally, we present the evaluation metrics.
\vspace{-6pt}
\subsection{Dataset}
The Librispeech 960-hours training set is used for training the acoustic module and the phonetic duration model \cite{panayotov2015librispeech}. 
TIMIT \cite{garofolo1993darpa} dataset is used to evaluate the phonetic duration model. 
The duration information in TIMIT is quantified to 30ms frames to align with the training set. There are phoneme mismatches between TIMIT and Librispeech. Therefore, for evaluation, we convert all the phonemes in TIMIT to Librispeech-style. 
For mispronunciation detection evaluation, a CAPT dataset with 2.8-hour English reading speech from 129 non-native children is collected. The dataset is annotated by five English teachers, including the phonetic and sentence-level annotation. The phonetic annotation follows the guideline in \cite{witt1999use}. For sentence-level annotation, the teachers are asked to evaluate the whole utterance with a score ranging from 0 to 10, regarding the utterance naturalness. In addition, 50\% of the data is used as the development set for hyper-parameter tuning, while the other 50\% is used for testing.
\vspace{-6pt}
\subsection{Implementation Details}

\noindent{\textbf{Acoustic Module and Decoding Module}}:
We employ the Hidden Markov Model-Time Delay Neural Network (HMM-TDNN) as the acoustic module. It is trained on Librispeech ASR task with Kaldi \cite{povey2011kaldi}. The input features are 40-dimensional Mel-Frequency Cepstrum Coefficients (MFCC) with 25-ms window size and 10-ms window shift. I-vectors are concatenated with MFCC to add some environment and speaker information. The HMM-TDNN applies discriminative chain settings with lattice-free maximum mutual information criterion \cite{povey2016purely}, sub-sampling techniques and the factorized mechanism \cite{peddinti2015time, povey2018semi}. Because of the down-sampling effect of sub-sampling TDNN model, the temporal resolution for the acoustic prediction is 30ms \cite{peddinti2015time}. For decoding, we align the phoneme sequence to the speech signal using the Viterbi algorithm (with optional silences between phonemes).

\vspace{5pt}
\noindent{\textbf{Duration Factor}}: Our proposed phonetic duration model uses 256-dimensional phoneme embeddings. For the encoder, we employ six four-head self-attention blocks with 256-dimension hidden states and 1024-dimension feed-forward layers. The output is frame-wise 1-dimensional, indicating the phonetic duration. The positional encoding follows the work in \cite{vaswani2017attention}. The parameters are initialized with uniform distribution as reported in \cite{glorot2010understanding}. Dropout is set as 0.1. Note that long phonetic sequences are constrained to a length of 100 for fast processing in training. 
We use L1 loss for training. We choose the Noam optimizer introduced in \cite{vaswani2017attention} with a learning rate of 0.001 and warm-up steps of 25,000. The batch size is set as 64. We use the clean validation set of Librispeech to select the best model for 100 epochs.

The baselines of the duration prediction includes a DNN-style model proposed in \cite{wei2019neural} and a LSTM model introduced in \cite{chen2017discrete}. The DNN performs a context expansion (seven left and seven right context). It uses six layers of 256 nodes, each with ReLU activation functions. The loss for the model is a cross-entropy loss. The LSTM models has three stacked LSTM layers with 256 nodes. We interpolate the cross-entropy loss with mean square error loss, as suggested in \cite{chen2017discrete}. In \cite{chen2017discrete, wei2019neural}, linguistic features are adopted such as position in word/phrase/sentences, Part of Speech, etc. However, to make the model comparable in terms of features, we only use the phoneme indicators and some phonetic properties (long/short vowels, voiced/unvoiced consonant, plosive, affricative, nasal, etc.). All the training configuration follows the self-attention model but with Adam optimizer.

\vspace{5pt}
\noindent{\textbf{Scoring module}}:
In the following experiments, the likelihood $p(o_t|a)$ and $p(o_1, o_2, ..., o_t|a)$ are computed using Viterbi approximation with the TDNN prediction and HMM alignment. For GOP and CaGOP, we determine the hyper-parameters (i.e., the phone-dependent thresholds for mispronunciation detection and $\beta$ in CaGOP) using the CAPT development set.
We compare five scoring modules in the experiments: 1) \textbf{GOP}; 2) \textbf{center GOP} which only uses the center frame of the segments for scoring; 3)  \textbf{CaGOP} which is our proposed scoring model; 4)  \textbf{CaGOP-TA} which removes the transition factor from CaGOP model; and 5) \textbf{CaGOP-Dur} which removes the duration factor from CaGOP model. From 4) and 5) we can analyze the contributions of the two factors in our proposed scoring module. 
\vspace{-10pt}
\subsection{Evaluation Metrics}
For duration prediction, we use the Mean Absolute Error (MAE) as the evaluation metrics. For mispronunciation detection, we adopt the accuracy and F1 measure. The labels (i.e., mispronounced or not) are imbalanced \footnote{In our case, more than 80\% phones are labeled as correct}, so F1 measure should be a more reliable metric. As stated in \cite{li2016mispronunciation}, long context of speech can serve as a better indicator of overall proficiency. Therefore, we also compute the sentence-level score by using the mean score of phonemes throughout the whole utterance. For sentence-level evaluation, we use Pearson correlation coefficients (PCC) and Spearman correlation coefficients (SCC), where the PCC focuses on numeric correlation, and SCC focuses on ranking correlation. The metric for the sentence-level score is the mean of correlation coefficients between each human rater and the evaluating scoring algorithm. 
\vspace{-10pt}
\section{Experimental Results}
Table \ref{tab:Duration Model} presents the frame-level phonetic duration prediction performance on the TIMIT dataset. It shows that our self-attention based duration model achieves 25\% relative improvement on MAE comparing to the best baseline LSTM system.

\vspace{-6pt}
\begin{table}[th]
  \caption{Frame-level Duration Prediction on TIMIT}
  \label{tab:Duration Model}
  \vspace{-4pt}
  \centering
  \begin{tabular}{ c  c }
    \toprule
    \textbf{Model} & \textbf{MAE (ms)} \\
    \midrule
     DNN \cite{henter2016robust} & 89.87 \\
     \hline
     LSTM \cite{chen2017discrete} & 42.74 \\
     \hline
     Self-attention  &   \textbf{31.88}   \\
    \bottomrule
  \end{tabular}
\end{table}
\vspace{-10pt}

The results of the mispronunciation detection task are shown in Table \ref{tab:mispron-acc}. The $\beta$ is set to 0.1 after tuned with the CAPT development set. Our proposed CaGOP significantly outperforms the GOP for more than 14\% absolute (20\% relative) improvement on both accuracy and F1. To be specific, the transition factor contributes 10\% absolute improvement while the duration factor contributes 4\% absolute improvement. Meanwhile, the center GOP, which only focuses on the center of the phonetic segment, gets a better result then GOP as well. It indicates that indeed there are phonetic transitions within the forced-alignment. After removing them from GOP, the scoring module can get better performance. Since the performance of `center GOP`  is worse than `$\text{CaGOP-Dur}$` which considers the transition factor, we can also conclude that the phonetic transitions are not always located around the boundary of the aligned segments. The result shows that our transition factor can successfully identify those "skewed" phonetic transitions.

The sentence-level scoring in Table \ref{tab:correlation} shows similar trend. When the human raters' PCC and SCC scores reach around 0.5, our CaGOP module achieves 0.392 PCC and 0.416 SCC. Compared to GOP baseline, CaGOP achieves 12\% relative improvement on both correlation metrics. Both the transition and duration factors contribute to the sentence-level scoring.

In addition to the above results, we also observe some interesting facts regarding the posterior-probability entropy of the acoustic module. First, the entropy of a phonetic segment is always in "U" shape. Second, the phonetic transitions (entropy peaks) depend on vocal tract shapes. For phones that are similar in vocal tract shapes, the transitions are not very significant (e.g. AH and N). Empirically, the "U" shapes are often right-skewed. We assume it might due to the acoustic module structure.

\vspace{-4pt}
\begin{table}[th]
  \caption{Model Performance on Mispronunciation Detection}
  \label{tab:mispron-acc}
  \vspace{-4pt}
  \centering
  \begin{tabular}{ c  c  c}
    \toprule
    \textbf{Method} & \textbf{Accuracy(\%)} & \textbf{F1(\%)}\\
    \midrule
     GOP \cite{witt1999use} & 53.44 & 65.81  \\
     \hline
     center GOP  & 60.88 & 72.98  \\
     \hline
     CaGOP  & \textbf{67.56}  & \textbf{79.89}   \\
      \hline
     CaGOP-Dur  &   64.07 & 75.85    \\
     \hline
     CaGOP-TA & 56.80 & 70.73 \\
    \bottomrule
  \end{tabular}
\end{table}
\vspace{-10pt}

\begin{table}[th]
  \caption{Correlation between Human-Rater \& Scoring Methods}
  \label{tab:correlation}
  \vspace{-4pt}
  \centering
  \begin{tabular}{ c  c  c  c}
    \toprule
    \textbf{Method} & \textbf{PCC} & \textbf{SCC}\\
    \midrule
     Human & 0.502 & 0.501 \\
     \hline
     GOP \cite{witt1999use} & 0.343 & 0.360 \\
     \hline
     center GOP  &   0.363 & 0.387   \\
      \hline
     CaGOP  &   \textbf{0.392} & \textbf{0.416}   \\
     \hline
     CaGOP-Dur  &   0.377 & 0.399    \\
     \hline
     CaGOP-TA & 0.375 & 0.393 \\
    \bottomrule
  \end{tabular}
\end{table}
\vspace{-10pt}



\section{Conclusion}
To deal with the limitations of GOP scoring, we propose the context-aware GOP (CaGOP) scoring model in this work, which injects two context related factors into the model, the transition factor and duration factor. The transition factor is represented using the frame-wise posterior-probability entropy. The duration factor is represented based on the duration mismatch, which is computed using a duration model with self-attention network. Experimental results prove the effectiveness of our proposed CaGOP scoring model, which achieves 20\% relative improvement at phoneme-level and 12\% relative improvement at the sentence-level over the GOP baselines. 
Both factors contribute to the performance boost. 
Our proposed duration model also achieves 25\% relative improvement for phonetic duration prediction on the TIMIT dataset.

\section{Acknowledgment}
This work was partially supported by National Natural Science Foundation of China (No. 61772535) and Beijing Natural Science Foundation (No. 4192028).

\bibliographystyle{IEEEtran}

\bibliography{mybib}

\begin{thebibliography}{10}
\providecommand{\url}[1]{#1}
\csname url@samestyle\endcsname
\providecommand{\newblock}{\relax}
\providecommand{\bibinfo}[2]{#2}
\providecommand{\BIBentrySTDinterwordspacing}{\spaceskip=0pt\relax}
\providecommand{\BIBentryALTinterwordstretchfactor}{4}
\providecommand{\BIBentryALTinterwordspacing}{\spaceskip=\fontdimen2\font plus
\BIBentryALTinterwordstretchfactor\fontdimen3\font minus
  \fontdimen4\font\relax}
\providecommand{\BIBforeignlanguage}[2]{{%
\expandafter\ifx\csname l@#1\endcsname\relax
\typeout{** WARNING: IEEEtran.bst: No hyphenation pattern has been}%
\typeout{** loaded for the language `#1'. Using the pattern for}%
\typeout{** the default language instead.}%
\else
\language=\csname l@#1\endcsname
\fi
#2}}
\providecommand{\BIBdecl}{\relax}
\BIBdecl

\bibitem{neri2002pedagogy}
A.~Neri, C.~Cucchiarini, H.~Strik, and L.~Boves, ``The pedagogy-technology
  interface in computer assisted pronunciation training,'' \emph{Computer
  assisted language learning}, vol.~15, no.~5, pp. 441--467, 2002.

\bibitem{evanini2013automated}
K.~Evanini and X.~Wang, ``Automated speech scoring for non-native middle school
  students with multiple task types.'' in \emph{Interspeech}, 2013, pp.
  2435--2439.

\bibitem{metallinou2014using}
A.~Metallinou and J.~Cheng, ``Using deep neural networks to improve proficiency
  assessment for children english language learners,'' in \emph{Fifteenth
  Annual Conference of the International Speech Communication Association},
  2014.

\bibitem{xu2009automatic}
S.~Xu, J.~Jiang, Z.~Chen, and B.~Xu, ``Automatic pronunciation error detection
  based on linguistic knowledge and pronunciation space,'' in \emph{2009 IEEE
  International Conference on Acoustics, Speech and Signal Processing}.\hskip
  1em plus 0.5em minus 0.4em\relax IEEE, 2009, pp. 4841--4844.

\bibitem{wang2015supervised}
Y.-B. Wang and L.-s. Lee, ``Supervised detection and unsupervised discovery of
  pronunciation error patterns for computer-assisted language learning,''
  \emph{IEEE/ACM Transactions on Audio, Speech, and Language Processing},
  vol.~23, no.~3, pp. 564--579, 2015.

\bibitem{witt2012automatic}
S.~M. Witt, ``Automatic error detection in pronunciation training: Where we are
  and where we need to go,'' in \emph{Proceedings of International Symposium on
  automatic detection on errors in pronunciation training}, vol.~1, 2012.

\bibitem{witt1997computer}
S.~Witt and S.~Young, ``Computer-assisted pronunciation teaching based on
  automatic speech recognition,'' \emph{Language Teaching and Language
  Technology Groningen, The Netherlands}, 1997.

\bibitem{tsubota2002recognition}
Y.~Tsubota, T.~Kawahara, and M.~Dantsuji, ``Recognition and verification of
  english by japanese students for computer-assisted language learning
  system,'' in \emph{Seventh International Conference on Spoken Language
  Processing}, 2002.

\bibitem{hu2013new}
W.~Hu, Y.~Qian, and F.~K. Soong, ``A new dnn-based high quality pronunciation
  evaluation for computer-aided language learning (call).'' in
  \emph{Interspeech}, 2013, pp. 1886--1890.

\bibitem{qian2012use}
X.~Qian, H.~Meng, and F.~K. Soong, ``The use of dbn-hmms for mispronunciation
  detection and diagnosis in l2 english to support computer-aided pronunciation
  training,'' in \emph{Thirteenth Annual Conference of the International Speech
  Communication Association}, 2012.

\bibitem{li2016mispronunciation}
K.~Li, X.~Qian, and H.~Meng, ``Mispronunciation detection and diagnosis in l2
  english speech using multidistribution deep neural networks,'' \emph{IEEE/ACM
  Transactions on Audio, Speech, and Language Processing}, vol.~25, no.~1, pp.
  193--207, 2016.

\bibitem{qian2017bidirectional}
Y.~Qian, K.~Evanini, X.~Wang, C.~M. Lee, and M.~Mulholland, ``Bidirectional
  lstm-rnn for improving automated assessment of non-native children's
  speech.'' in \emph{Interspeech}, 2017, pp. 1417--1421.

\bibitem{van2010using}
J.~v. Doremalen, C.~Cucchiarini, and H.~Strik, ``Using non-native error
  patterns to improve pronunciation verification,'' in \emph{Eleventh Annual
  Conference of the International Speech Communication Association}, 2010.

\bibitem{zhang2012exploit}
L.~Zhang, H.~Li, and L.~Ma, ``Exploit posterior probability algorithm for
  pronunciation quality evaluation,'' \emph{Journal of Computational
  Information Systems}, vol.~8, pp. 9251--9258, 11 2012.

\bibitem{fant2012auditory}
G.~Fant, \emph{Auditory analysis and perception of speech}.\hskip 1em plus
  0.5em minus 0.4em\relax Elsevier, 2012.

\bibitem{shannon2001mathematical}
C.~E. Shannon, ``A mathematical theory of communication,'' \emph{ACM SIGMOBILE
  mobile computing and communications review}, vol.~5, no.~1, pp. 3--55, 2001.

\bibitem{panayotov2015librispeech}
V.~Panayotov, G.~Chen, D.~Povey, and S.~Khudanpur, ``Librispeech: an asr corpus
  based on public domain audio books,'' in \emph{2015 IEEE International
  Conference on Acoustics, Speech and Signal Processing (ICASSP)}.\hskip 1em
  plus 0.5em minus 0.4em\relax IEEE, 2015, pp. 5206--5210.

\bibitem{dahl2011context}
G.~E. Dahl, D.~Yu, L.~Deng, and A.~Acero, ``Context-dependent pre-trained deep
  neural networks for large-vocabulary speech recognition,'' \emph{IEEE
  Transactions on audio, speech, and language processing}, vol.~20, no.~1, pp.
  30--42, 2011.

\bibitem{tong2016context}
R.~Tong, N.~F. Chen, B.~Ma, and H.~Li, ``Context aware mispronunciation
  detection for mandarin pronunciation training.'' in \emph{Interspeech}, 2016,
  pp. 3112--3116.

\bibitem{zeleny2012linear}
M.~Zeleny, \emph{Linear multiobjective programming}.\hskip 1em plus 0.5em minus
  0.4em\relax Springer Science \& Business Media, 2012, vol.~95.

\bibitem{pylkkonen2004duration}
J.~Pylkkonen and M.~Kurimo, ``Duration modeling techniques for continuous
  speech recognition,'' in \emph{Eighth International Conference on Spoken
  Language Processing}, 2004.

\bibitem{tokuda2016temporal}
K.~Tokuda, K.~Hashimoto, K.~Oura, and Y.~Nankaku, ``Temporal modeling in neural
  network based statistical parametric speech synthesis.'' in \emph{SSW}, 2016,
  pp. 106--111.

\bibitem{henter2016robust}
G.~E. Henter, S.~Ronanki, O.~Watts, M.~Wester, Z.~Wu, and S.~King, ``Robust tts
  duration modelling using dnns,'' in \emph{2016 IEEE International Conference
  on Acoustics, Speech and Signal Processing (ICASSP)}.\hskip 1em plus 0.5em
  minus 0.4em\relax IEEE, 2016, pp. 5130--5134.

\bibitem{chen2017discrete}
B.~Chen, T.~Bian, and K.~Yu, ``Discrete duration model for speech synthesis.''
  in \emph{Interspeech}, 2017, pp. 789--793.

\bibitem{wei2019neural}
X.~Wei, M.~Hunt, and A.~Skilling, ``Neural network-based modeling of phonetic
  durations,'' in \emph{Interspeech}, 2019, pp. 1751--1755.

\bibitem{vaswani2017attention}
A.~Vaswani, N.~Shazeer, N.~Parmar, J.~Uszkoreit, L.~Jones, A.~N. Gomez,
  {\L}.~Kaiser, and I.~Polosukhin, ``Attention is all you need,'' in
  \emph{Advances in neural information processing systems}, 2017, pp.
  5998--6008.

\bibitem{sperber2018self}
M.~Sperber, J.~Niehues, G.~Neubig, S.~St{\"u}ker, and A.~Waibel,
  ``Self-attentional acoustic models,'' in \emph{Interspeech}, 2018, pp.
  3723--3727.

\bibitem{garofolo1993darpa}
J.~S. Garofolo, L.~F. Lamel, W.~M. Fisher, J.~G. Fiscus, and D.~S. Pallett,
  ``Darpa timit acoustic-phonetic continous speech corpus cd-rom. nist speech
  disc 1-1.1,'' \emph{NASA STI/Recon technical report n}, vol.~93, 1993.

\bibitem{witt1999use}
S.~M. Witt \emph{et~al.}, ``Use of speech recognition in computer-assisted
  language learning,'' Ph.D. dissertation, University of Cambridge Cambridge,
  United Kingdom, 1999.

\bibitem{povey2011kaldi}
D.~Povey, A.~Ghoshal, G.~Boulianne, L.~Burget, O.~Glembek, N.~Goel,
  M.~Hannemann, P.~Motlicek, Y.~Qian, P.~Schwarz \emph{et~al.}, ``The kaldi
  speech recognition toolkit,'' in \emph{IEEE 2011 workshop on automatic speech
  recognition and understanding}, no. CONF.\hskip 1em plus 0.5em minus
  0.4em\relax IEEE Signal Processing Society, 2011.

\bibitem{povey2016purely}
D.~Povey, V.~Peddinti, D.~Galvez, P.~Ghahremani, V.~Manohar, X.~Na, Y.~Wang,
  and S.~Khudanpur, ``Purely sequence-trained neural networks for asr based on
  lattice-free mmi.'' in \emph{Interspeech}, 2016, pp. 2751--2755.

\bibitem{peddinti2015time}
V.~Peddinti, D.~Povey, and S.~Khudanpur, ``A time delay neural network
  architecture for efficient modeling of long temporal contexts,'' in
  \emph{Sixteenth Annual Conference of the International Speech Communication
  Association}, 2015.

\bibitem{povey2018semi}
D.~Povey, G.~Cheng, Y.~Wang, K.~Li, H.~Xu, M.~Yarmohammadi, and S.~Khudanpur,
  ``Semi-orthogonal low-rank matrix factorization for deep neural networks.''
  in \emph{Interspeech}, 2018, pp. 3743--3747.

\bibitem{glorot2010understanding}
X.~Glorot and Y.~Bengio, ``Understanding the difficulty of training deep
  feedforward neural networks,'' in \emph{Proceedings of the thirteenth
  international conference on artificial intelligence and statistics}, 2010,
  pp. 249--256.

\end{thebibliography}


\end{document}